\documentclass[aps,twocolumn,groupedaddress,nofootinbib,amsmath,amssymb]{revtex4}
\usepackage{dsfont}
\usepackage{bm}
\begin{document}

\title{Spinning black holes for generalized scalar tensor theories in three dimensions}

\author{Olaf Baake}
\email{olaf.baake-at-inst-mat.utalca.cl} \affiliation{Instituto de
Matem\'atica y F\'isica, Universidad de Talca, Casilla 747 Talca,
Chile}
\affiliation{Centro de Estudios Cientificos (CECs), Av. Arturo Part 514, Valdivia, Chile.}
\author{Mois\'es~Bravo-Gaete}
\email{ mbravo-at-ucm.cl} \affiliation{Facultad de Ciencias
B\'asicas, Universidad Cat\'olica del Maule, Casilla 617, Talca, Chile.}

\author{Mokhtar~Hassaine}
\email{hassaine-at-inst-mat.utalca.cl} \affiliation{Instituto de
Matem\'atica y F\'isica, Universidad de Talca, Casilla 747 Talca,
Chile}

\begin{abstract}
We consider a general class of scalar tensor theories in three dimensions whose action contains up to second-order derivatives of the scalar field with coupling functions that only depend on the standard kinetic term of the scalar field, thus ensuring the
invariance under the constant shift of the scalar field. For this model, we
show that the field equations for a stationary metric ansatz together with a purely radial scalar field can be fully integrated. The kinetic term of the scalar field solution is shown to satisfy an algebraic relation depending only on the coupling functions, and
hence is constant while the metric solution is nothing but the BTZ metric with an effective cosmological constant fixed in terms of the coupling functions. As a direct consequence the thermodynamics of the solution is shown to be identical to the BTZ one with an effective cosmological constant, despite the presence of a scalar field. Finally, the expression of the semi-classical entropy of this
solution is also confirmed through a generalized Cardy-like formula involving the mass of the scalar soliton obtained from the black hole by means of a double Wick rotation.
\end{abstract}

\maketitle

%%%%%%%%%%%%%%%%%%%%%
\section{Introduction}
%%%%%%%%%%%%%%%%%%%%%

Since the discovery of the BTZ black hole solution
\cite{Banados:1992wn}, the study of three-dimensional gravity has received considerable attention to such an extent that it is now considered an interesting laboratory to explore the many facets of the lower-dimensional physics at the classical level but also at the quantum level. By three-dimensional gravity we are referring not only to Einstein's
standard action but to all of its possible variations, including, for
example, its higher-order massive theories, such as the Topologically
Massive Gravity \cite{Deser:1982vy}, or the New Massive Gravity
\cite{Bergshoeff:2009hq}. The three-dimensional gravity models, with
or without matter source, are likewise of importance due to the variety of their solutions, and particularly their asymptotic AdS black hole solutions whose near horizon geometry can
be relevant to test some conceptual aspects of the AdS/CFT correspondence \cite{Maldacena:1997re, Brown:1986nw}. In this aspect, the BTZ solution is of particular interest because its in-depth study over the past three decades has considerably enhanced our knowledge on the statistical interpretation of the black hole entropy, see e. g. \cite{Strominger:1997eq, Carlip:2007ph, Frodden:2012nu}. It is further fascinating that BTZ-like metrics arise as solutions of radically different three-dimensional gravity models. To illustrate this statement, we could mention for
example the emergence of BTZ-like solutions in the context of
massive gravity \cite{Hendi:2016pvx}, in higher-order theories
\cite{Konoplya:2020ibi}, but also in the presence of matter source, such
as a scalar (dilatonic) field, see e. g. \cite{Chan:1994qa, Hendi:2017oka}. In the present work, we will confirm this trend by showing that the equations of motion of a general class of scalar tensor theories, enjoying a shift symmetry of the scalar field, and
involving up to second-order derivatives of the scalar field, can be fully integrated and solved by the BTZ metric.

The interests of studying scalar tensor theories is mainly due to the fact that it constitutes one of the simplest modified gravity theories by extending General Relativity with one or more scalar degree of freedom. The dedication to scalar theories is not new and
its origin may be attributed a posteriori to the seminal work of Horndeski \cite{Horndeski:1974wa}, who presented the most general scalar tensor theory in four dimensions with second order equations of motion. The requirement not to have more than two derivatives in
the equations of motion is connected to the Ostrogradski theorem which states that (under certain assumptions) higher-order derivative theories have a Hamiltonian that is unbounded from below. This is related to the appearance of an extra (ghost) degree of freedom with negative energy. Thus the absence of higher time derivatives in the equations of motion guarantees the absence of the Ostrogradski ghost. Nevertheless, it has been shown recently that some particular higher-order theories of a single scalar field extension of General Relativity can propagate healthy degrees of freedom and are mechanically stable. The most general such Lagrangian depending quadratically on second-order derivatives of a scalar field was constructed in \cite{Motohashi:2016ftl, BenAchour:2016fzp}, and dubbed Degenerate Higher Order Scalar Tensor (DHOST) theory. This terminology indicates that the absence of Ostrogradski ghosts is mainly due to the degeneracy property of its Lagrangian.
There even exists a subclass of DHOST theories where gravitational waves propagate at the speed of light, being in perfect agreement with the observed results \cite{Langlois:2017dyl}. While these attractive properties of scalar tensor theories occur in four dimensions, we nevertheless like to explore the implications of such models in three dimensions. This is precisely the aim of the present work.

Here, we will consider a general scalar tensor theory in three dimensions with a field content given by the metric $g$ and a scalar field denoted by $\phi$. The main assumption concerning the action is its invariance under the constant translation of the scalar
field, i. e. $\phi\to \phi+\mbox{const.}$ which implies the existence of a conserved Noether charge. It is known that this hypothesis considerably simplifies the integration of the equations of motion. The action will contain up to second-order covariant
derivatives of the scalar field and is parity invariant, that is invariant under the discrete transformation $\phi\to -\phi$. The action is parameterized in terms of six coupling functions that depend only on the kinetic term $X=g^{\mu\nu}\partial_{\mu}\phi\partial_{\nu}\phi$ of the scalar field. Recently it has been shown that such scalar tensor theories are invariant under a Kerr-Schild symmetry, and this symmetry turns out to be extremely useful for generating black hole solutions from simple seed configurations \cite{Babichev:2020qpr}. Here we will adopt a different strategy by deriving the most general stationary solution by brute force, as it was done for the special case of
Horndeski theory in three dimensions \cite{Babichev:2020qpr}. Interestingly enough, we will show that the integration of the equations of motion forces the scalar field to have a constant kinetic term while at the same time the metric functions turn out to be the BTZ spacetime with an effective cosmological constant expressed in terms of the coupling functions appearing in the action. We would like to emphasize that the constant value of the kinetic scalar field term results from an algebraic equation that $X$ must
satisfy, and consequently it does not correspond to any hair. Although the metric solution is given by the BTZ metric, it is legitimate to wonder whether the presence of the scalar field could affect the thermodynamic properties of the solution. In order to answer this question, the thermodynamics of the solution is carefully analyzed within the Euclidean method \cite{Gibbons:1976ue, Regge:1974zd}, and it is shown that the expressions of the mass, entropy and angular momentum are identical to those of the BTZ solution with an effective cosmological constant. In addition, since it has been pointed out that the Wald formula for the entropy \cite{Wald:1993nt} applied to general scalar tensor theories may be
problematic \cite{Feng:2015oea}, we have found it sensible to compute the entropy of the solution by means of a generalized Cardy formula. In this formulation the ground state is identified with a soliton whose mass is proportional to the lowest eigenvalues of the shifted Virasoro operators, see \cite{Correa:2010hf, Gonzalez:2011nz, Correa:2012rc, Ayon-Beato:2019kmz, BravoGaete:2017dso}. In order to achieve this task, we have constructed the static scalar soliton from the black hole through a double Wick rotation and computed its mass. Finally, the application of the generalized Cardy formula is shown to properly reproduce the semi-classical expression of the entropy.

The plan of the paper is organized as follows. In the next section we will present the action and derive the most general solution for a stationary ansatz for the metric together with a radial scalar field. We will show that the metric solution is nothing other than
the BTZ metric while the kinetic term of the scalar field is constant. In Section III we will construct the regularized Euclidean action which allows us to identify the mass, the angular momentum and the entropy. Further, the expression of the entropy will be
confirmed through a computation involving the generalized Cardy formula and the mass of the static scalar soliton. The mass of the soliton will be computed using the quasi local formalism \cite{Kim:2013zha}. Finally, in Section IV we present our conclusions and discussions.

%%%%%%%%%%%%%%%%%%%%%%%%%%%%%%%%%%%%%%%%%%%%%%%%%%%%%%%%%%%%%%%
\section{Scalar field model and the derivation of its solution}
%%%%%%%%%%%%%%%%%%%%%%%%%%%%%%%%%%%%%%%%%%%%%%%%%%%%%%%%%%%%%%%
In three dimensions, we are considering a scalar tensor theory whose dynamical fields are represented by a metric, $g$, and a scalar field ,$\phi$. The action reads {\small
\begin{eqnarray}
S &=&\int
d^3x\sqrt{-g}{\cal{L}}\nonumber\\
&=&\int
d^3x\sqrt{-g}\Big[Z(X)+G(X)R+A_3(X)\Box\phi\,\phi^{\mu}\phi_{\mu\nu}\phi^{\nu}\nonumber\\
&+&A_2(X)\left((\Box\phi)^2-\phi_{\mu\nu}\phi^{\mu\nu}\right)+A_4(X)\phi^{\mu}\phi_{\mu\nu}\phi^{\nu\rho}\phi_{\rho}\nonumber\\
&+&A_5(X)\left(\phi^{\mu}\phi_{\mu\nu}\phi^{\nu}\right)^2\Big],
\label{action2}
\end{eqnarray}}
where for simplicity we have defined $X=\partial_{\mu}\phi\,\partial^{\mu}\phi$ and
$\phi_{\mu\nu}=\nabla_{\mu}\nabla_{\nu}\phi$. Here, the six coupling functions $Z, G$ and $A_i$ for $i=2,\cdots 5$ are a priori arbitrary functions of the kinetic term $X$, and contain up to second-order covariant derivatives of the scalar field. It is easy to see that the action is invariant under the shift symmetry $\phi\to \phi+\mbox{const.}$, as well as under the discrete transformation $\phi\to -\phi$. The field equations of the action (\ref{action2}) are reported in the Appendix.

We now look for black hole solutions with a stationary metric and a purely radial scalar field. The most general such ansatz can be parametrized as follows
\begin{eqnarray}
&& ds^2=-f(r)dt^2+\frac{dr^2}{f(r)}+H^2(r)\left[d\theta-k(r)dt\right]^2,\nonumber\\
&& \phi=\phi(r). \label{ansatz2}
\end{eqnarray}
After some tedious computations one can show that the field equations associated to the action (see Appendix) will become fully integrable for the ansatz (\ref{ansatz2}) by fixing the coupling function $A_5$ in terms of the others through the following relation
\begin{eqnarray}
A_5=\frac{\left(2A_2+XA_3+4G_X\right)^2}{2X(G+XA_2)}-\frac{A_3+A_4}{X},
\label{A5}
\end{eqnarray}
where $G_X=\frac{dG}{dX}$. This relation is quite similar to the four-dimensional DHOST conditions which ensure the absence of Ostrogradski ghosts \cite{Motohashi:2016ftl, BenAchour:2016fzp}. Further, the emergence of the condition (\ref{A5}) is not surprising,
since in the literature concerning scalar tensor theories of the type (\ref{action2}), most of the solutions are found for special relations between the coupling functions $Z, G$ and the $A_i$'s, see e.g. \cite{Babichev:2017guv,Chagoya:2018lmv,Babichev:2017lmw,Kobayashi:2018xvr,Ruth,Takahashi:2019oxz,Minamitsuji:2019tet,Lehebel:2018zga}.

In what follows, we will consider the action (\ref{action2}) with the coupling function $A_5$ given by the relation (\ref{A5}), and for later convenience, we also define the following expressions
\begin{subequations}
\label{defZ}
\begin{eqnarray}
\mathcal{Z}_1 &=& G + X A_2, \label{defZ1}\\
\mathcal{Z}_2 &=& 2 A_2 + X A_3 + 4 G_X. \label{defZ2}
\end{eqnarray}
\end{subequations}

We are now in the position to present the general derivation of the spinning solution. As a first step, we consider the following combination of the metric equations
\begin{eqnarray*}
\mathcal{E}_{t\theta} + k \mathcal{E}_{\theta\theta} = 0,
\end{eqnarray*}
which yields a first integral given by
\begin{eqnarray}
\Big( \mathcal{Z}_1 H^3 k' \Big)' = 0. \label{kprime}
\end{eqnarray}
Further, the combination
\begin{eqnarray*}
2 \sqrt{f^3 X} \mathcal{E}_{rr} - J^r = 0,
\end{eqnarray*}
where $J^r$ is the radial component of the conserved current (see Appendix) permits to express the derivative of the metric function $f$ as
{\small
\begin{eqnarray}
f' = -\frac{4 f H' \mathcal{Z}_1 \mathcal{Z}_2 X'+f H
\mathcal{Z}_2^2 X'^2+4 H^3 k'^2 \mathcal{Z}_1^2-8 H Z
\mathcal{Z}_1}{8 H' \mathcal{Z}_1^2+2 H \mathcal{Z}_1 \mathcal{Z}_2
X'}. \label{fprime}
\end{eqnarray}}
Inserting this into the two combinations
\begin{subequations}
\begin{eqnarray}
\frac{\mathcal{Z}_2}{f} \left( \mathcal{E}_{tt} + k \mathcal{E}_{t\theta} \right) &=& 0, \\
\frac{\mathcal{Z}_1}{\sqrt{f X}} \mathcal{E}_{J} +
\frac{\mathcal{Z}_2}{H^2} \mathcal{E}_{t\theta} &=& 0,
\end{eqnarray}
\end{subequations}
one obtains after some manipulations the following equation
\begin{eqnarray}
 k \mathcal{Z}_{2} \Big( \mathcal{Z}_1 H^3 k' \Big)' +4 H \left[\left( \mathcal{Z}_1 Z \right)_X - Z \mathcal{Z}_2\right]   = 0.
\end{eqnarray}
By equation (\ref{kprime}) the first term vanishes, leaving
\begin{eqnarray}
\left( \mathcal{Z}_1 Z \right)_X - Z \mathcal{Z}_2 = 0.
\label{algeX}
\end{eqnarray}
It is easy to see that, for $Z=0$, equation (\ref{algeX}) is completely degenerate and gives no information about the kinetic term. Consequently in what follows we impose $Z$ to be nonzero. Then the kinetic term $X$ must satisfy this algebraic equation which in turn implies that $X$ has to be constant. Moreover, we would like to stress that its constant value is not an integration constant but must be rather understood as follows: Given a scalar tensor theory (\ref{action2}-\ref{A5}) with specific coupling functions $Z, G$ and the $A_i$'s, the constant value of $X$ will be determined by the algebraic relation (\ref{algeX}). \\
This restriction on the kinetic term of the scalar field significantly simplifies the equations, in particular the combination
\begin{eqnarray}
\frac{2}{f} \left( \mathcal{E}_{tt} + 2 k \mathcal{E}_{t\theta} + k
\mathcal{E}_{\theta\theta} \right) + 2 f \mathcal{E}_{rr} +
\sqrt{\frac{X}{f}} J^r = 0
\end{eqnarray}
implies
\begin{eqnarray}
H'' = 0. \label{Hsecond}
\end{eqnarray}
Finally, after some redefinitions of the coordinates, the metric solution can be casted in the standard BTZ form as {\small
\begin{subequations}
\label{bhsol}
\begin{eqnarray}
\label{ansatzsol}
ds^2 = - N(r)^2F(r)dt^2 + \frac{dr^2}{F(r)}
+r^2 \left(d
\theta+N^{\theta}(r) dt\right)^2,\label{metricbtz}\\
N=1,\, F=\left( \frac{Z}{2 \mathcal{Z}_1} r^2 -M+ \frac{J^2}{4
r^2}\right),\, N^{\theta}=\frac{J}{2 r^2}.\qquad \label{solbtz}
\end{eqnarray}
\end{subequations}}
This metric is nothing but the BTZ solution with an effective cosmological cosmological constant given by $\Lambda_{{\tiny\mbox{eff}}}=-Z/2 \mathcal{Z}_1$. Various comments can be made concerning the emergence of the BTZ metric solution together with a scalar field with constant kinetic term. First of all, it is remarkable that the equations of motion of the general class of scalar tensor theories, given by the action (\ref{action2}) together with the condition (\ref{A5}), is fully integrable and yield the BTZ solution. It is also remarkable that, although the model is defined in terms of  the coupling functions $Z, G$ and $A_2, A_3$ and $A_4$, the resulting solution is shown to be parametrized in
terms of $Z$ and through the combinations $\mathcal{Z}_1$ and $\mathcal{Z}_2$, as defined in Eq. (\ref{defZ}). This would also imply that scalar tensor theories of the form (\ref{action2}-\ref{A5}) with different coupling functions $Z, G$ and $A_i$'s can have the
same effective cosmological constant $\Lambda_{{\tiny\mbox{eff}}}$, and hence they can be solved by the same BTZ metric. Moreover, it is easy to see that the action (\ref{action2}-\ref{A5}) enjoys a Kerr-Schild symmetry as defined in \cite{Babichev:2020qpr} whose implementation on the stationary ansatz (\ref{ansatz2}) can be summarized as
\begin{eqnarray}
f(r)\to f(r)-a(r),\,\, H(r)\to H(r),\,\,k(r)\to k(r),
\end{eqnarray}
with a constant mass term i.e. $a(r)=M$ which is a direct consequence of the kinetic term $X$ being constant. This can be put in analogy with the four-dimensional static case where solutions of the action (\ref{action2}) with constant kinetic term were shown to have the standard Coulomb mass term $a(r)=\frac{M}{r}$, see Ref. \cite{Babichev:2020qpr}.

In summary, we have shown that for a stationary ansatz (\ref{ansatz2}) the integration of the field equations of (\ref{action2}-\ref{A5}) forces $X$ to be constant together with a BTZ metric (\ref{metricbtz})-(\ref{solbtz}). In the following section, we analyze its thermodynamics.

%%%%%%%%%%%%%%%%%%%%%%%%%%%%%%%%%%%%%%%%%%%%%%%%%%%
\section{Thermodynamics of the spinning solution}
%%%%%%%%%%%%%%%%%%%%%%%%%%%%%%%%%%%%%%%%%%%%%%%%%%%%%%%

The thermodynamics of the solution will now be determined by means of the Euclidean method \cite{Gibbons:1976ue, Regge:1974zd}, where the Euclidean continuation of the metric is obtained by setting $t=-i\tau$ in the ansatz (\ref{ansatzsol}). In order for the resulting metric to be real, one can introduce a complex constant of integration for the Euclidean momentum as $J_{\tiny{\mbox{Eucl}}}=-iJ$, where $J$ will be identified with the physical angular momentum. In order to avoid a conical singularity, the Euclidean time $\tau$ has to be made periodic with period $\beta=1/T$, where $T$ is the temperature that in our case is given by
\begin{eqnarray}\label{hawkingtemp}
T=\frac{F'(r)}{4 \pi} \Big{|}_{r=r_{h}}=\frac{1}{4\pi} \left(\frac{2
r_h}{L^2}-\frac{J^2}{2 r_h^3}\right),
\end{eqnarray}
and where for simplicity we have defined the square of the effective AdS radius
$$
L^2=\frac{2 \mathcal{Z}_1(X)}{Z(X)}.
$$
Recall that the kinetic term $X$ is a constant determined by the algebraic relation defined by Eq. (\ref{algeX}). After some computations, the Euclidean action is shown to be given by
\begin{eqnarray*}
I_{E}&=&2 \pi \beta \int_{r_h}^{+\infty}dr \Big\{N\Big[F'\left(\frac{1}{4} r \big(F (\phi')^2\big)' \mathcal{Z}_2+\mathcal{Z}_1\right)\\
&+&\frac{1}{2} F \left(4 \big(F (\phi')^2\big)'\mathcal{Z}_{1X}+r \mathcal{Z}_2 \big(F (\phi')^2\big)''\right)\nonumber\\
&+&\frac{1}{4} F r \Big(\big(F (\phi')^2\big)'\Big)^2 \left(2
\mathcal{Z}_{2X}-\frac{\mathcal{Z}_2^2}{2 \mathcal{Z}_{1}}\right)
-Z r\nonumber\\
&+&\frac{1}{2} \frac{p^2}{\mathcal{Z}_1r^3}\Big]+N^{\theta}
p'\Big\}+B_{E},
\end{eqnarray*}
where $r_h$ is the radius of the event horizon and
\begin{eqnarray*}
p(r)=\frac{r^3 \big(N^{\theta}\big)' \mathcal{Z}_{1}}{N}.
\end{eqnarray*}
The Euclidean action $I_E$ is defined up to a boundary term $B_{E}$ which is fixed such that said action has an extremum, that is $\delta I_{E}=0$. In the present case, the variation of this boundary term can be conveniently expressed as
\begin{eqnarray*}
&&\delta B_E=-2\pi\beta \left[\left(\left(\frac{\delta I_{E}}{\delta F'}\right)-
\left(\frac{\delta I_{E}}{\delta F''}\right)'\right)\delta F  \right.\\
&&\left.+\left(\frac{\delta I_{E}}
{\delta F''}\right) \delta F'+\left(\frac{\delta I_{E}}{\delta \phi''}\right) \delta \phi' +\left(\frac{\delta I_{E}}{\delta \phi'''}\right)
\delta \phi''\right.\\
&&\left. -\left(\frac{\delta I_{E}}{\delta \phi'''}\right)' \delta \phi'+\left(\left(\frac{\delta I_{E}}{\delta \phi'}\right)-\left(\frac{\delta I_{E}}{\delta \phi''}\right)'
 \right.\right.\\
&&\left.\left.+\left(\frac{\delta I_{E}}{\delta \phi'''}\right)''+2 F \phi' \left(\frac{\delta I_{E}}{\delta
X}\right)\right) \delta \phi +N^{\theta} \delta p\right]_{r=r_h}^{r=+\infty}.
\end{eqnarray*}
At infinity, most of these terms cancel each other out, yielding
\begin{eqnarray*}
\delta B_{E} \Big{|}_{+\infty}=2 \pi \beta \mathcal{Z}_{1} \delta M
\Rightarrow B_{E}\big{|}_{+\infty}= 2 \pi \beta \mathcal{Z}_{1} M,
\end{eqnarray*}
while that at the horizon
\begin{eqnarray*}
\delta B_{E} \Big{|}_{r_h}&=&8\mathcal{Z}_{1}\pi^2 \delta r_h- 2\pi\beta \Omega\, \delta(\mathcal{Z}_{1} J)\Rightarrow \\
B_{E}\big{|}_{r_h}&=& 8\mathcal{Z}_{1}\pi^2 r_h-2\pi \beta \Omega\,
\mathcal{Z}_{1} J.
\end{eqnarray*}
In this expression, $\Omega$ represents the chemical potential, defined by
$$
\Omega=\lim _{r \rightarrow +\infty}N^{\theta}(r)-N^{\theta}(r_h)=-\frac{J}{2 r_h^2}.
$$

With all of the above, the boundary term $B_{E}$ is simply expressed as
\begin{eqnarray}
B_{E}&=&B_{E}\big{|}_{+\infty}-B_{E}\big{|}_{r_h}\nonumber \\
&=&2 \pi \beta \mathcal{Z}_{1} M-8\mathcal{Z}_{1}\pi^2 r_h+2\pi\beta
\Omega\,\mathcal{Z}_{1} J. \label{Bound}
\end{eqnarray}
Finally, the thermodynamic quantities can be read off from the Gibbs free energy $F$
\begin{equation}\label{free_ene}
I_{E}=\beta F = \beta{\cal{M}}-{\cal{S}}-\beta \Omega {\cal{J}},
\end{equation}
where ${\cal{M}}$ is the mass, ${\cal{S}}$ the entropy and, as before,  $\Omega$ is the chemical potential associated with the angular momentum ${\cal{J}}$, see \cite{Gibbons:1976ue}. Finally, comparing (\ref{Bound}) with (\ref{free_ene}), the thermodynamic parameters
turn out to be given by
\begin{subequations}
\label{thermoqtes}
\begin{eqnarray}
{\cal{S}}&=&8\mathcal{Z}_{1}\pi^2 r_h,\label{entropy}\\
{\cal{M}}&=&2 \pi \mathcal{Z}_{1} M=2 \pi \mathcal{Z}_{1}\left(\frac{r_h^2}{L^2}+{\frac {{J}^2}{4 r_h^2}}\right),\label{mass}\\
{\cal{J}}&=&-2 \pi \mathcal{Z}_{1} J,\qquad  \Omega=-\frac{J}{2
r_h^2},\label{momentum-vel}
\end{eqnarray}
\end{subequations}
and one can easily see that the first law holds, namely $d {\cal{M}}=T d{\cal{S}}+\Omega d{\cal{J}}$. These thermodynamic quantities (\ref{thermoqtes}) are identical to those of the BTZ solution with an effective AdS radius given by $L$.

We now proceed by re-deriving the expression of the semi-classical
entropy (\ref{entropy}) by means of a generalized Cardy formula. In this formulation, the entropy of the black hole solution can be microscopically computed provided the theory admits a regular scalar soliton which would be identified as the ground state of the theory, see \cite{Correa:2010hf, Correa:2012rc}. In our case, the regular soliton will be obtained from the static black hole solution (\ref{bhsol}) with $J=0$ through a double Wick rotation $t \rightarrow i \theta \mbox{ and } \theta \rightarrow i t$ together with a identification for the location of the event horizon $r_h=L$ given by
\begin{eqnarray*}
ds^2=-\frac{r^2}{L^2} dt^2+\left(\frac{r^2}{L^2}-1\right)^{-1} dr^2+\left(\frac{r^2}{L^2}-1\right)d \theta^2,\nonumber \\
\end{eqnarray*}
and the line element of the regular static scalar solution after a redefinition of the radial coordinate reads
\begin{eqnarray}
ds^2=-L^4 \cosh^2(\rho)dt^2+L^2 d \rho^2+L^2 \sinh^2(\rho)d
\theta^2. \label{solads}
\end{eqnarray}
As done for example in Refs. \cite{Ayon-Beato:2019kmz,BravoGaete:2017dso}, the mass of the soliton (\ref{solads}) will be computed within the quasi local formalism defined in \cite{Kim:2013zha}. In order to be as self-contained as possible, we will elaborate the steps of the
computations. To begin with, the variation of the action (\ref{action2}-\ref{A5}) can be schematically represented as
\begin{eqnarray}
\delta S= \sqrt{-g}\left[\varepsilon_{\mu \nu} \delta g^{\mu
\nu}+\varepsilon_{(\phi)} \delta \phi\right]+\partial_{\mu}
\Theta^{\mu} (\delta g, \delta \phi),
\end{eqnarray}
where $\varepsilon_{\mu \nu}$ and $\varepsilon_{(\phi)}$ corresponds
to the equations of motions with respect to the metric $g_{\mu \nu}$
and the scalar field $\phi$ (see the Appendix), while $\Theta^{\mu}$ is
a surface term whose expression is given by
\begin{eqnarray*}
\Theta^{\mu}&=&\sqrt{-g}\Big[2\left(P^{\mu (\alpha\beta)
\gamma}\nabla_{\gamma}\delta g_{\alpha\beta}-\delta g_{\alpha\beta} \nabla_{\gamma}P^{\mu(\alpha\beta)\gamma}\right)\\
&+&\frac{\delta \cal{L}}{\delta (\phi_{\mu})} \delta
\phi-\nabla_{\nu}\left(\frac{\delta \cal{L}}{\delta (\phi_{\mu
\nu})}\right) \delta \phi
+\frac{\delta \cal{L}}{\delta (\phi_{\mu \nu})} \delta (\phi_{\nu})\\
&-&\frac{1}{2}\frac{\delta \cal{L}}{\delta (\phi_{\mu \sigma})}
\phi^{\sigma} \,
\delta g_{\sigma \rho}-\frac{1}{2}\frac{\delta \cal{L}}{\delta (\phi_{\sigma \mu})}\phi^{\sigma} \,\delta g_{\sigma \rho}\\
&+&\frac{1}{2}\frac{\delta \cal{L}}{\delta ( \phi_{\sigma
\rho})}\phi^{\mu}\,\delta g_{\sigma \rho}\Big],
\end{eqnarray*}
with $P^{\mu\nu\lambda\rho}=\delta {\cal{L}}/ \delta R_{\mu\nu\lambda\rho}$, and ${\cal{L}}$ is the Lagrangian. Considering now the variation induced by a diffeomorphism generated by a Killing vector $\xi^{\mu}$ whose action on the metric and the scalar field read
\begin{eqnarray*}
\delta_{\xi} g_{\mu \nu}&=&2 \nabla_{(\mu} \xi_{\nu)},\qquad \delta_{\xi} \phi=\xi^{\sigma} (\nabla_{\sigma} \phi),\\
\delta_{\xi} (\nabla_{\nu}
\phi)&=&\xi^{\sigma}\phi_{\sigma \nu}+(\nabla_{\nu}\xi^{\sigma}) \phi_{\sigma},
\end{eqnarray*}
we construct a Noether current given by
$${\cal{L}}\,\xi^{\mu}+2 \varepsilon^{\mu \nu} \xi_{\nu}-\Theta^{\mu}(\delta_{\xi} g,\delta_{\xi} \phi)=\nabla_{\nu} K^{\mu \nu},$$
which is derived from the potential  $K^{\mu\nu}$,
\begin{eqnarray*}
K^{\mu\nu}&=&\sqrt{-g}\,\Big[2P^{\mu\nu\rho\sigma}\nabla_\rho \xi_\sigma -4\xi_\sigma
\nabla_\rho P^{\mu\nu\rho\sigma}+\frac{\delta \cal{L}}{ \delta \phi_{\mu \sigma}} \phi^{\nu} \xi_{\sigma}\\
&-&\frac{\delta \cal{L}}{ \delta \phi_{\nu \sigma}} \phi^{\mu}
\xi_{\sigma}\Big].
\end{eqnarray*}
As shown in \cite{Kim:2013zha}, for each Killing field a corresponding conserved quantity can be constructed as
\begin{equation}
\label{chargequasi} Q(\xi)\!=\!\int_{\cal B}\!
dx_{\mu\nu}\Big(\delta K^{\mu\nu}(\xi)-2\xi^{[\mu} \!\! \int^1_0ds~
\Theta^{\nu]}\Big),
\end{equation}
Here, $\delta K^{\mu\nu}(\xi) =
K^{\mu\nu}_{s=1}(\xi)-K^{\mu\nu}_{s=0}(\xi)$ is the difference of the Noether potential interpolating between the solutions along the path parametrized by $s\in [0,1]$, and $dx_{\mu\nu}$ represents the integration over the two-dimensional boundary ${\cal B}$. For the Killing field, $\xi=\partial_t$, one obtains that $$
\delta K^{r t}= -\frac{2 G}{L} \qquad \int_{0}^{1} ds
\Theta^{r}=-\frac{\mathcal{Z}_{1}}{L}+\frac{2 G}{L},
$$
yielding a mass for the static soliton (\ref{solads}) that given by
\begin{equation}\label{masssoliton}
{\cal{M}}_{\tiny{\mbox{sol}}}=-2 \pi \mathcal{Z}_{1}.
\end{equation}

We are now in position to provide a microscopic computation of the black hole entropy. As stressed in \cite{Correa:2010hf}, the Cardy formula is more conveniently expressed in terms of the vacuum charge rather than the central charge:
\begin{eqnarray}
\label{Cardyf} {\cal{S}}_{C}=4\pi\sqrt{-\tilde{\Delta}^+_0\tilde{\Delta}^+}+
4\pi\sqrt{-\tilde{\Delta}^-_0\tilde{\Delta}^-},
\end{eqnarray}
where ($\tilde{\Delta}^{\pm}_0$) $\tilde{\Delta}^{\pm}$ are the (lowest) eigenvalues of the shifted Virasoro operators. The eigenvalues are related to the mass and angular momentum as
\cite{Strominger:1997eq}
$$
{\cal
M}=\frac{1}{L}\left(\tilde{\Delta}^{+}+\tilde{\Delta}^{-}\right),\qquad
{\cal J}=\tilde{\Delta}^{+}-\tilde{\Delta}^{-}.
$$
On the other hand, since the scalar soliton is identified with the ground state of the theory, its mass (\ref{masssoliton}) is proportional to the lowest eigenvalue
$$
\tilde{\Delta}^{\pm}_0=\frac{L}{2}{\cal{M}}_{\tiny{\mbox{sol}}}.
$$
Finally, the Cardy formula (\ref{Cardyf}) can be conveniently rewritten in terms of ${\cal M}, {\cal J}$ as
\begin{eqnarray*}
{\cal{S}}_{C}=2\pi \sqrt{-L{\cal{M}}_{\tiny{\mbox{sol}}}\left(L {\cal M}+{\cal
J}\right)}+2\pi \sqrt{-L {\cal{M}}_{\tiny{\mbox{sol}}}\left(L{\cal
M}-{\cal J}\right)},
\end{eqnarray*}
and it can be verified that this correctly reproduces the semi-classical entropy (\ref{entropy}), this is, ${\cal{S}}_{C}={\cal{S}}$.

%%%%%%%%%%%%%%%%%%%%
\section{Conclusions and discussions}
%%%%%%%%%%%%%%%%%%%%

In the present work, we have shown that the equations of motion of a very general class of scalar tensor theories (\ref{action2}-\ref{A5})
can be fully integrated for a stationary metric ansatz together with a purely radial scalar field. Interestingly enough, the kinetic
term of the scalar field solution was forced to be constant, while at the same time the spacetime metric resulted to be the BTZ metric
with an effective cosmological constant expressed in terms of the coupling functions. It is somehow appealing that the spectrum of such
general class of theories only consists of the BTZ metric with (different) effective
cosmological constants. This observation is even more relevant considering that in four dimensions, theories which are much less general
than that studied here admit black hole solutions that are asymptotically AdS, flat or even exhibit a rather exotic asymptotic
behavior
\cite{Babichev:2017guv,Chagoya:2018lmv,Babichev:2017lmw,Kobayashi:2018xvr,Ruth,Takahashi:2019oxz,Minamitsuji:2019tet,Lehebel:2018zga}. Even more, in four dimensions a recipe has even been given to construct
black hole solutions from any simple seed metric \cite{Babichev:2020qpr}. Nevertheless, one can notice an important difference
concerning the kinetic term of the scalar field solution between the three and the four dimensional
situations. Indeed, solutions in four dimensions with non-constant kinetic term were shown to
exist \cite{Minamitsuji:2019tet, Babichev:2020qpr}, while in our case the algebraic relation (\ref{algeX})
forces the kinetic term to be constant. One might also have
thought that the presence of a coupled scalar field should have affected the thermodynamics of the solution
but this was not the case. This is essentially due to the constancy of the kinetic term of the scalar field solution.
It would be nice to provide a physical explanation for the emergence of the BTZ metric as the solution of such a very
general class of scalar tensor theories (\ref{action2}-\ref{A5}).

It is further intriguing that the equations of motion become fully
integrable by imposing the condition (\ref{A5}) on the coupling
function $A_5$. As mentioned before, this relation is quite similar
to the four dimensional DHOST conditions \cite{Motohashi:2016ftl,
BenAchour:2016fzp} which prevent the emergence of Ostrogradski
ghosts. It would be compelling to explore this point more deeply.
Moreover, it is worth mentioning that said BTZ solution remains a
solution even if one replaces the scalar field ansatz with $\phi = q
t + \psi(r) + L \theta$ in (\ref{ansatz2}), and if $X$ still solves
the algebraic equation (\ref{algeX}). Note that in this case, the
vanishing of the radial component of the current $J^r=0$ is a
consequence of the field equation \cite{Babichev:2015rva}. In
\cite{Babichev:2020qpr} the uniqueness of this solution was shown
for the quadratic Horndeski action. Whether or not it is unique in
the general case has yet to be established.

In \cite{Babichev:2020qpr}, it was shown that the solution generating method also applies
for generalized Proca theories with solutions having a non-zero radial component for the potential \cite{Chagoya:2016aar}. Hence, in a complete analogy
with the work done here, it will be interesting to look for black hole solutions in three dimensions for more general vector tensor theories \cite{Heisenberg:2017hwb}.

%%%%%%%%%%%%%%%%%%%%%%%%%%%
\section*{Acknowledgements}
%%%%%%%%%%%%%%%%%%%%%%%%%%%
OB is funded by the PhD scholarship of the University of Talca. MH
gratefully acknowledges the kind support of the ECOSud project
C18U04. MB is supported by grant Programa Fondecyt
de Iniciaci\'on en Investigaci\'on No. 11170037.

%%%%%%%%%%%%%%%%%%
\section{Appendix}
\label{sec:app}
%%%%%%%%%%%%%%%%%%
Here we report the equations of motion of the action (\ref{action2}) that are obtained by varying the action with respect to the metric  $\mathcal{E}_{\mu\nu}$ and those with respect to the scalar field $\varepsilon_{(\phi)}$. The former are given by
\begin{eqnarray}
\mathcal{E}_{\mu\nu}:={\cal{G}}^{Z}_{\mu\nu}+{\cal{G}}^{G}_{\mu\nu}+\sum_{i=2}^{5}{\cal{G}}^{(i)}_{\mu\nu}=0,
\end{eqnarray}
where
\begin{align*}
{\cal{G}}^{Z}_{\mu\nu}&=-\frac{1}{2} Z(X) g_{\mu\nu}+K_{X} \phi_{\mu}
\phi_{\nu}, \\ \\
{\cal{G}}^{G}_{\mu\nu}&= G  G_{\mu\nu}+G_{X} R \phi_{\mu} \phi_{\nu} -\nabla_{\nu} \nabla_{\mu} G  \\
&+g_{\mu\nu} \nabla_{\lambda}\nabla^{\lambda} G,\\ \\
{\cal{G}}^{(2)}_{\mu\nu}&=-\phi_{\mu}\,(A_{2X} \nabla_{\nu}X)\,\Box \phi-(A_{2X} \nabla_{\mu} X ) \phi_{\nu} \,\Box \phi\\
&-A_2  \phi_{\nu \mu} \Box \phi -\phi_{\nu \mu} \phi_{\lambda} ( A_{2X} \nabla^{\lambda}
X)
\\
&+\phi_{\nu} \phi_{\lambda \mu} (A_{2X} \nabla^{\lambda}X)+\phi_{\mu}
\phi_{\lambda\nu} (A_{2X} \nabla^{\lambda} X)\\
&+A_2  R_{\nu\lambda} \phi_{\mu} \phi^{\lambda}+A_2  R_{\mu\lambda} \phi_{\nu}
\phi^{\lambda}\\
&-A_2  \phi_{\lambda\nu\mu} \phi^{\lambda}+\frac{1}{2}A_2  g_{\mu\nu} (\Box \phi)^2\\
&+g_{\mu\nu} \phi_{\lambda}  (A_{2X} \nabla^{\lambda} X) \Box \phi +A_2  g_{\mu\nu}
\phi^{\lambda}  \phi^{\,\,\rho \,}_{\rho \,\,\lambda}
\\
&-A_2  g_{\mu\nu} R_{\lambda\rho} \phi^{\lambda} \phi^{\rho}
+\frac{1}{2} A_2  g_{\mu\nu} \phi_{\rho\lambda} \phi^{\rho\lambda}\\
&+A_{2X} \phi_{\mu} \phi_{\nu}  \big((\Box \phi)^2-\phi_{\lambda\rho}
\phi^{\lambda\rho}\big),\\ \\
{\cal{G}}^{(3)}_{\mu\nu}&=-\frac{1}{2}A_3  \phi_{\mu} \phi_{\nu} (\Box
\phi)^2-
\frac{1}{2} \phi_{\mu} \phi_{\nu} \phi_{\lambda}  (A_{3X} \nabla^{\lambda}X) \Box \phi\\
&+\frac{1}{2} A_{3}  \phi_{\mu}  \phi_{\lambda\nu} \phi^{\lambda} \Box \phi+
\frac{1}{2} A_{3}  \phi_{\nu} \phi_{\lambda\mu}\phi^{\lambda}
\Box \phi\\
&-\frac{1}{2} A_{3}  \phi_{\mu} \phi_{\nu} \phi^{\lambda} \phi_{\rho\,\, \lambda}^{\,\,
\rho \,} +\frac{1}{2} A_3  R_{\lambda\rho} \phi_{\mu} \phi_{\nu} \phi^{\lambda}
\phi^{\rho}\\
&-\frac{1}{2} \phi_{\mu}  (A_{3X} \nabla_{\nu} X) \phi^{\lambda}  \phi_{\rho\lambda}
\phi^{\rho} -\frac{1}{2} (A_{3X} \nabla_{\mu} X )\phi_{\nu}
\phi^{\lambda} \phi_{\rho\lambda} \phi^{\rho} \\
&-\frac{1}{2}A_3  \phi_{\nu} \phi^{\lambda} \phi_{\rho\lambda\mu}
\phi^{\rho}-\frac{1}{2}A_3
\phi_{\mu} \phi^{\lambda}  \phi_{\rho\lambda\nu} \phi^{\rho}\\
&-A_3  \phi_{\nu} \phi^{\lambda} \phi_{\rho\lambda} \phi^{\rho}_{\,\,\mu}-A_3  \phi_{\mu}
\phi^{\lambda}
\phi_{\rho\lambda} \phi^{\rho}_{\,\ \nu} \\
&+\frac{1}{2} g_{\mu\nu} \phi_{\lambda} (A_{3X} \nabla^{\lambda} X) \phi^{\rho}
\phi_{\sigma\rho}  \phi^{\sigma} +\frac{1}{2} g_{\mu\nu} A_{3}  \phi^{\lambda}  \phi^{\rho}
\phi_{\sigma\rho\lambda} \phi^{\sigma}\\
&+g_{\mu\nu} A_{3}  \phi^{\lambda}  \phi^{\rho} \phi_{\sigma\rho}
\phi^{\sigma}_{\,\,\lambda}+A_{3X} \phi_{\mu}
\phi_{\nu} (\Box \phi) \phi^{\rho} \phi_{\sigma\rho} \phi^{\sigma}, \\ \\
{\cal{G}}^{(4)}_{\mu\nu}&=-A_4  \phi_{\mu} \phi_{\nu} \phi^{\lambda}
\phi_{\rho\,\,\lambda}^{\,\,\rho}+ A_4  \phi_{\lambda\mu} \phi^{\lambda} \phi_{\rho\nu} \phi^{\rho}
\\
&-\phi_{\mu}  \phi_{\nu}  (A_{4X} \nabla^{\lambda} X) \phi_{\rho\lambda} \phi^{\rho} -A_4
\phi_{\mu} \phi_{\nu}
\phi_{\rho\lambda} \phi^{\rho\lambda}\\
&-\frac{1}{2} A_4  g_{\mu\nu} \phi^{\lambda} \phi^{\rho} \phi_{\sigma\rho}
\phi^{\sigma}_{\,\,\lambda} +A_{4X} \phi_{\mu} \phi_{\nu} \phi_{\lambda\rho} \phi^{\lambda}
\phi^{\rho\sigma} \phi_{\sigma},\\ \\
{\cal{G}}^{(5)}_{\mu\nu}&=-A_5  \phi_{\mu} \phi_{\nu} \phi^{\lambda} \phi_{\rho\lambda}
\phi^{\rho} (\Box \phi)
-\phi_{\mu} \phi_{\nu}  \phi_{\lambda} (A_{5X} \nabla^{\lambda} X) \phi^{\rho} \phi_{\sigma\rho} \phi^{\sigma}\\
&+A_5  \phi_{\nu}  \phi_{\lambda\mu} \phi^{\lambda} \phi^{\rho} \phi_{\sigma\rho} \phi^{\sigma}+
A_5  \phi_{\mu} \phi_{\lambda\nu} \phi^{\lambda} \phi^{\rho}
\phi_{\sigma\rho} \phi^{\sigma}\\
&-A_5  \phi_{\mu} \phi_{\nu}  \phi^{\lambda} \phi^{\rho}
\phi_{\sigma\rho\lambda} \phi^{\sigma}-2A_5  \phi_{\mu} \phi_{\nu}  \phi^{\lambda} \phi^{\rho} \phi_{\sigma\rho} \phi^{\sigma}_{\,\,\lambda}\\
&-\frac{1}{2} A_5  g_{\mu\nu} \phi^{\lambda} \phi_{\rho\lambda} \phi^{\rho} \phi^{\sigma}
\phi_{\tau\sigma} \phi^{\tau}+A_{5X} \phi_{\mu}  \phi_{\nu}
 \phi^{\lambda} \phi^{\rho} \phi_{\rho\lambda}
\phi^{\sigma}\phi^{\tau} \phi_{\tau\sigma},
\end{align*}
while the field equations associated to the scalar field allow to construct a current conservation equation given by
\begin{eqnarray*}
\varepsilon_{(\phi)}=\nabla_{\mu}
J^{\mu}=\nabla_{\mu}\left[\frac{\delta \cal{L}}{\delta
(\phi_{\mu})}-\nabla_{\nu}\left(\frac{\delta \cal{L}}{\delta (
\phi_{\mu \nu})}\right)\right]=0,
\end{eqnarray*}
where
\begin{eqnarray*}
J^{\mu}=J^{\mu}_{Z}+J^{\mu}_{G}+\sum_{i=2}^{5} J^{\mu}_{(i)},
\end{eqnarray*}
with
\begin{eqnarray*}
J^{\mu}_{Z}&=&2 Z_{X} \phi^{\mu}, \\ \\
J^{\mu}_{G}&=&2 G_{X} R \phi^{\mu}, \\ \\
J^{\mu}_{(2)}&=& 2 A_{2X} \phi^{\mu}\left[ (\Box \phi)^{2}-
\phi_{\lambda \rho}\phi^{\lambda \rho}\right]-2\nabla_{\nu} \left[ A_2  \left( g^{\mu \nu}-\phi^{\mu \nu}\right)\right], \\ \\
J^{\mu}_{(3)}&=&2 A_{3X} \phi^{\mu}  \, \Box \phi\,  \phi^{\lambda}
\phi_{\lambda \rho} \phi^{\rho}+2 A_3  \,\Box \phi \,
\phi^{\mu}_{\,\,\,\lambda}\phi^{\lambda}\\
&-&\nabla_{\nu} \left[A_3  \big(g^{\mu \nu} \phi^{\lambda} \phi_{\lambda \rho} \phi^{\rho}+\Box \phi \,\phi^{\mu} \phi^{\nu}\big)\right],
\end{eqnarray*}
\begin{eqnarray*}
J^{\mu}_{(4)}&=&2 A_{4X} \phi^{\mu} \phi^{\sigma} \phi_{\sigma \rho}
\phi^{\rho \lambda}
\phi_{\lambda}+A_4(X) \big[\phi^{\mu}_{\,\,\rho} \phi^{\rho \lambda} \phi_{\lambda}\\
&+&\phi^{\sigma} \phi_{\sigma \rho} \phi^{\rho \mu} \big]-\nabla_{\nu}\big[A_4(X) \big(\phi^{\mu} \phi^{\nu \rho} \phi_{\rho}\\
&+&\phi^{\sigma} \phi_{\,\,\,\sigma}^{\mu}
\phi^{\nu}\big)\big], \\ \\
J^{\mu}_{(5)}&=&2 A_{5X} \phi^{\mu} \big( \phi^{\sigma} \phi_{\sigma
\rho} \phi^{\rho}\big)^2 +2 A_5(X)\big(
\phi^{\sigma} \phi_{\sigma \rho} \phi^{\rho}\big)\big(\phi^{\mu \sigma}\phi_{\sigma}\\
&+&\phi^{\sigma \mu} \phi_{\sigma}\big)-2\nabla_{\nu}\left[A_5(X)
\phi^{\sigma} \phi_{\sigma \rho} \phi^{\rho}  \phi^{\mu}
\phi^{\nu}\right].
\end{eqnarray*}

%%%%%%%%%%%%%%%%%%%%%%%%%%%%%%%%%

\end{document}